\documentclass{aa} 

\usepackage{epsfig}     
\usepackage{graphicx}     
\usepackage{amssymb}            
\usepackage{url}                
\usepackage{amsmath}            
\usepackage{rotating}                   
\usepackage{float}                      
\usepackage{textcomp}           
\usepackage{epstopdf}
\usepackage{dcolumn}
\usepackage{txfonts}
\usepackage{tabularx}
\usepackage{hyperref}
\hypersetup{
    colorlinks,
    citecolor=blue,
    filecolor=blue,
    linkcolor=blue,
    urlcolor=blue}
\usepackage{soul} 
\usepackage[english]{babel}
\usepackage{booktabs}
\usepackage{gensymb}
\usepackage{subfig}
\usepackage{appendix}

\usepackage{etoolbox}

\begin{document}

\title{Evidence of external reconnection between an erupting mini-filament and ambient loops observed by Solar Orbiter/EUI}

\author{Z. F. Li\inst{1,2,3}, X. Cheng\inst{1,2,3}\fnmsep\thanks{Corresponding author: Xin Cheng \email{xincheng@nju.edu.cn}}, M. D. Ding\inst{1,2}, L. P. Chitta\inst{3}, H. Peter\inst{3}, D. Berghmans\inst{4}, P. J. Smith\inst{5}, F. Auch\`{e}re\inst{6}, S. Parenti\inst{6}, K. Barczynski\inst{7,8}, L. Harra\inst{7,8}, U. Sch\"{u}hle\inst{3}, \'{E}. Buchlin\inst{6}, C. Verbeeck\inst{4}, R. Aznar Cuadrado\inst{3}, A. N. Zhukov\inst{4,9}, D. M. Long\inst{5}, L. Teriaca\inst{3}, L. Rodriguez\inst{4}}

\institute{School of Astronomy and Space Science, Nanjing University, Nanjing, 210046, People's Republic of China
\and Key Laboratory of Modern Astronomy and Astrophysics (Nanjing University), Ministry of Education, Nanjing 210093, China
\and Max Planck Institute for Solar System Research, Justus-von-Liebig-Weg 3, 37077 G{\"o}ttingen, Germany
\and Solar-Terrestrial Centre of Excellence – SIDC, Royal Observatory of Belgium, Ringlaan -3- Av. Circulaire, 1180 Brussels, Belgium
\and UCL-Mullard Space Science Laboratory, Holmbury St. Mary, Dorking, Surrey RH5 6NT, UK
\and Universit\'{e} Paris-Saclay, CNRS, Institut d'Astrophysique Spatiale, 91405 Orsay, France
\and Physikalisch-Meteorologisches Observatorium Davos, World Radiation Center, 7260 Davos Dorf, Switzerland
\and ETH-Z{\"u}rich, Wolfgang-Pauli-Str. 27, 8093 Z{\"u}rich, Switzerland
\and Skobeltsyn Institute of Nuclear Physics, Moscow State University, 119992 Moscow, Russia}

\date{received ...; accepted ...}

\abstract
{Mini-filament eruptions are one of the most common small-scale transients in the solar atmosphere. However, their eruption mechanisms are still not understood thoroughly. Here, with a combination of 174 \AA\,images of high spatio-temporal resolution taken by the Extreme Ultraviolet Imager on board Solar Orbiter and images of the Atmospheric Imaging Assembly on board Solar Dynamics Observatory, we investigate in detail an erupting mini-filament over a weak magnetic field region on 2022 March 4. Two bright ribbons clearly appeared underneath the erupting mini-filament as it quickly ascended, and subsequently, some dark materials blew out when the erupting mini-filament interacted with the outer ambient loops, thus forming a blowout jet characterized by a widening spire. At the same time, multiple small bright blobs of 1--2 Mm appeared at the interaction region and propagated along the post-eruption loops toward the footpoints of the erupting fluxes at a speed of $\sim$ 100 km s$^{-1}$. They also caused a semi-circular brightening structure. Based on these features, we suggest that the mini-filament eruption first experiences internal and then external reconnection, the latter of which mainly transfers mass and magnetic flux of the erupting mini-filament to the ambient corona.}

\keywords{Sun: flares; Sun: magnetic fields; Sun: corona}
\titlerunning{Evidence of external reconnection}
\authorrunning{Li et al.}
\maketitle

\begin{figure*}[!htb]
    \centering
    \includegraphics[width=0.85\textwidth,clip,trim=0cm 0cm 0cm 0cm]{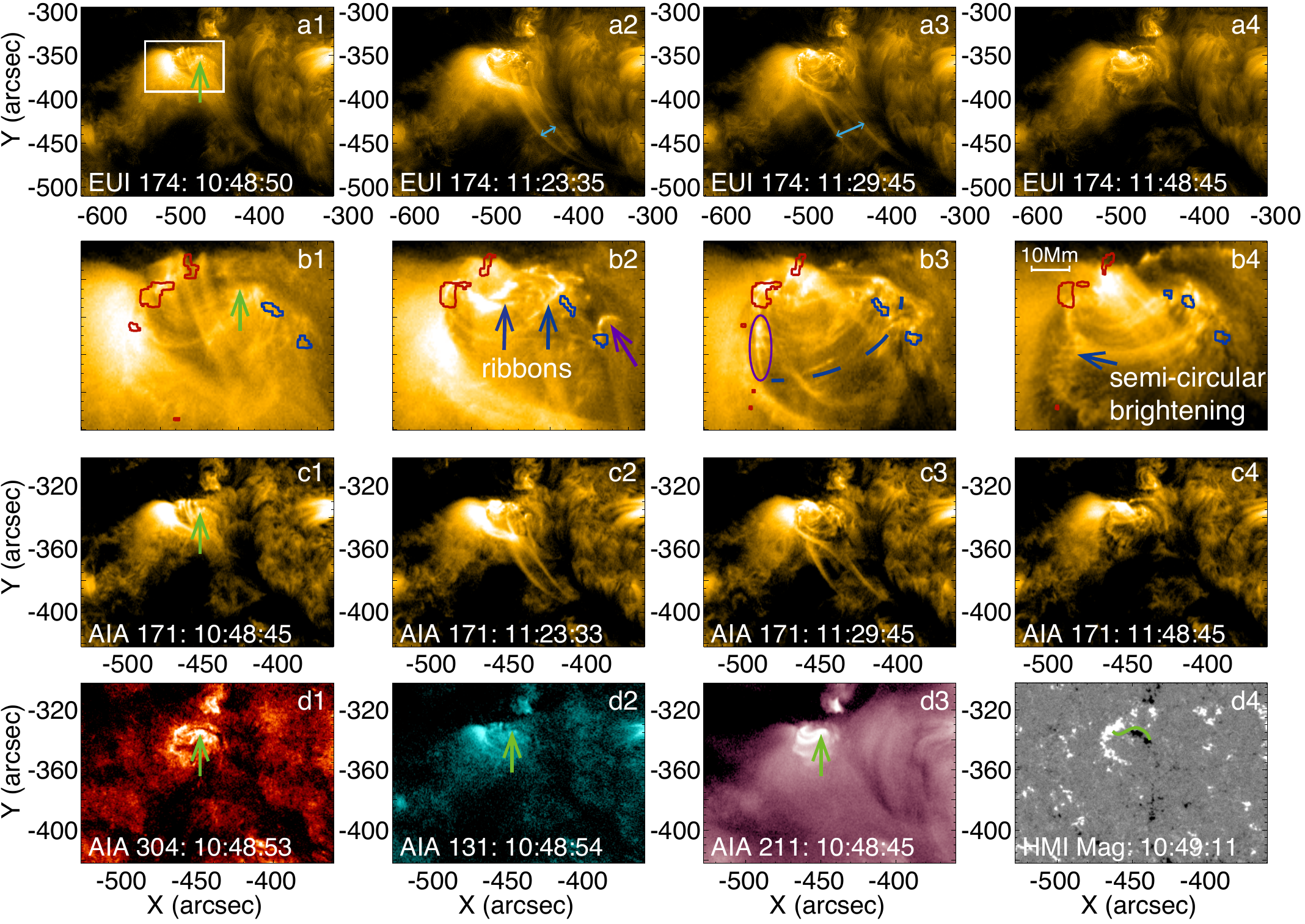}
    \caption{Overview of observations. a. Time sequence of HRI$_{EUV}$ 174 \AA\ images showing the temporal evolution of the mini-filament eruption. The light blue arrows in panels a2 and a3 show the widths of the ambient loops. b. Zoom-in of the mini-filament eruption (white box in panel a1) overlaid with the contours ($\pm$ 40 G) of the HMI LOS magnetogram, with the positive (negative) in red (blue). The bright ribbons and semi-circular brightening (arrows in dark blue) are pointed out in panels b2 and b4, respectively. The arrow in purple indicates the brightening at the mini-filament footpoint. The dashed dark blue line denotes the post-eruption loops in panel b3. The purple ellipse in panel b3 highlights a plasma blob that is analyzed in detail in Fig. \ref{fig4}d. c. AIA 171 \AA\ images showing the mini-filament eruption. d. AIA 304 \AA, 131 \AA, 211 \AA\ images, and HMI magnetogram at the moment before the eruption. The locations of the mini-filament are indicated by the arrows in green. The green curve in panel d4 represents the basic shape of the mini-filament. An animation for the evolution of the mini-filament eruption and its zoom-in at 174 \AA\  is available online.} 
    \label{fig1}
\end{figure*}

\section{Introduction}
\label{intro}

Mini-filaments are small-scale analogs to solar filaments. As observed against the solar disk in chromospheric H$\alpha$ images, they extend less than 25 Mm and last for about 50 minutes \citep{Wang2000}. Mini-filament eruptions occur with a high frequency of about 6000 per day over the whole solar quiescent regions \citep{Wang2000,Sakajiri2004}. Similar to large-scale filaments, mini-filaments usually lie above the photospheric polarity inversion line (PIL) of small-scale adjacent opposite-polarity magnetic fields. The eruptions of mini-filaments are also found to often produce mini coronal mass ejections and microflares \citep{Innes2009}.

Many mini-filament eruptions are associated with coronal jets that represent transient, collimated plasma ejections along open fields or far-reaching coronal loops \citep{Raouafi2016}. They have been observed primarily at soft X-ray (SXR; \citealt{Shibata1992,Alexander1999}), extreme ultraviolet (EUV; \citealt{Wang1998}), and H$\alpha$ wavelengths (known as ``surges"; \citealt{Chae1999,Canfield1996}), reflecting the fact that jets may have both hot and cool components. \citet{Moore2010} introduced the concept of blowout jets for the first time and found that they tend to have wider spires and more dynamic bases than standard jets \citep{Shibata1992}. Subsequently, several studies found that small-scale filament eruptions tend to produce a blowout jet under open field conditions such as coronal holes or edges of active regions via external reconnection \citep{Shen2012,Adams2014,Li2015}. Recently, a so-called mini-filament eruption model has been put forward, in which a successful mini-filament eruption drives a blowout jet, while a partial or failed mini-filament eruption causes a standard jet \citep{Sterling2015,Sterling2022}. In addition to the external reconnection, this model also involves internal reconnection beneath the erupting mini-filaments, which would form heated arcades, namely the jet bright points (JBPs; they are thought to correspond to the solar flare arcades in the larger-scale cases), appearing near or at the locations below the erupting mini-filaments.

In the past years, many features of blowout jets were constantly revealed, including spinning motions in the spire \citep{Jiang2007}, lateral expansion \citep{Shen2011}, and the formation of plasmoids or blobs \citep{Zhang2017}. In particular, the blobs have drawn much more attention as they provide strong evidence of the magnetic reconnection process during blowout jets. So far, these blobs have been observed in many cases. \citet{Singh2012} observed multiple bright plasma ejections in chromospheric anemone jets. \citet{Zhang2014} reported the blobs in recurrent and homologous EUV jets for the first time. However, in their study, only one blob was observed for each jet. The blobs have also been observed in a blowout surge by \citet{Li2017}. \citet{Huang2018} even found that moving bright blobs transferred mass between coronal loops and a nearby arch filament system through magnetic reconnection. A recent study using data from Solar Orbiter \citep{Muller2020} with unprecedented high resolution found that the blobs were mainly restricted along the jet base \citep{Mandal2022}. \citet{Long2023} also observed small blobs that drained down the legs of the loop system. In spite of these progresses, it is still difficult to distinctly clarify the relation between the moving plasma blobs and magnetic reconnection, in particular during blowout jets. The clarification is mainly limited by the low spatio-temporal resolution of previous data. 

Taking advantage of extremely high spatial and temporal resolution data of the the Extreme Ultraviolet Imager \cite[EUI;][]{Rochus2020} on board Solar Orbiter, we observed multiple plasma blobs originating from the interaction region of the erupting mini-filament and the ambient loops that propagated along the post-eruption loops, thus showing evidence of external reconnection between the erupting and ambient flux. Section \ref{data} describes the data we used. The results are presented in Sect. \ref{results}, which is followed by a summary and discussion in Sect. \ref{discussion}. 

\section{Data}
\label{data}

We primarily used imaging data of the quiescent Sun near the disk center taken with the EUI on board Solar Orbiter. In particular, we used data of the High Resolution Imager at 174 \AA\ (HRI$_{EUV}$), which samples plasma with a temperature T of $\sim$ 1 MK. The EUI data\footnote{https://doi.org/10.24414/2qfw-tr95} were obtained on 2022 March 4 between 10:48 -- 11:48 UT when Solar Orbiter was located 0.54 AU away from the Sun and was about 7 degrees east in solar longitude from the Sun-Earth line, with a pixel size of 0.492$^{\prime\prime}$ and a time cadence of 5 s \citep{FirstPerihelionofEUI}. At the time of these observations, one pixel size corresponds to approximately 190 km on the Sun.

We also used the EUV images from the Atmospheric Imaging Assembly (AIA; \citealt{Lemen2012}) on board the Solar Dynamics
Observatory \cite[SDO;][]{Pesnell2012} from seven EUV passbands, including 304 \AA, 171 \AA, 193 \AA, 211 \AA, 131 \AA, 94~\AA, and 335 \AA, with a time cadence of 12 s, a pixel size of 0.6$^{\prime\prime}$, and a field of view of 1.3 $R_{\odot}$. The projected distance of one pixel on the Sun of AIA is $\sim$ 400 km. The AIA data are processed with the $aia\_prep.pro$ and deconvolved with $aia\_deconvolve\_richardsonlucy$ routines, both of which are available in the SolarSoft Ware (SSW) package \citep{Freeland1998}. The alignment of the AIA and HRI$_{EUV}$ images is achieved through a cross correlation between the AIA 171 \AA\,and HRI$_{EUV}$ 174 \AA\,images with similar fields of view. Our magnetic field data are from full-disk line-of-sight magnetograms of SDO/Helioseismic and Magnetic Imager \cite[HMI;][]{Scherrer2012}, with a cadence of 45 s and a pixel size of 0.5$^{\prime \prime}$. The time of the EUI images was adapted to that at 1 AU, taking different distances from the Sun between Solar Orbiter and SDO into account.

\begin{figure}
    \centering
    \includegraphics[width=0.50\textwidth,clip,trim=0cm 0cm 0cm 0cm]{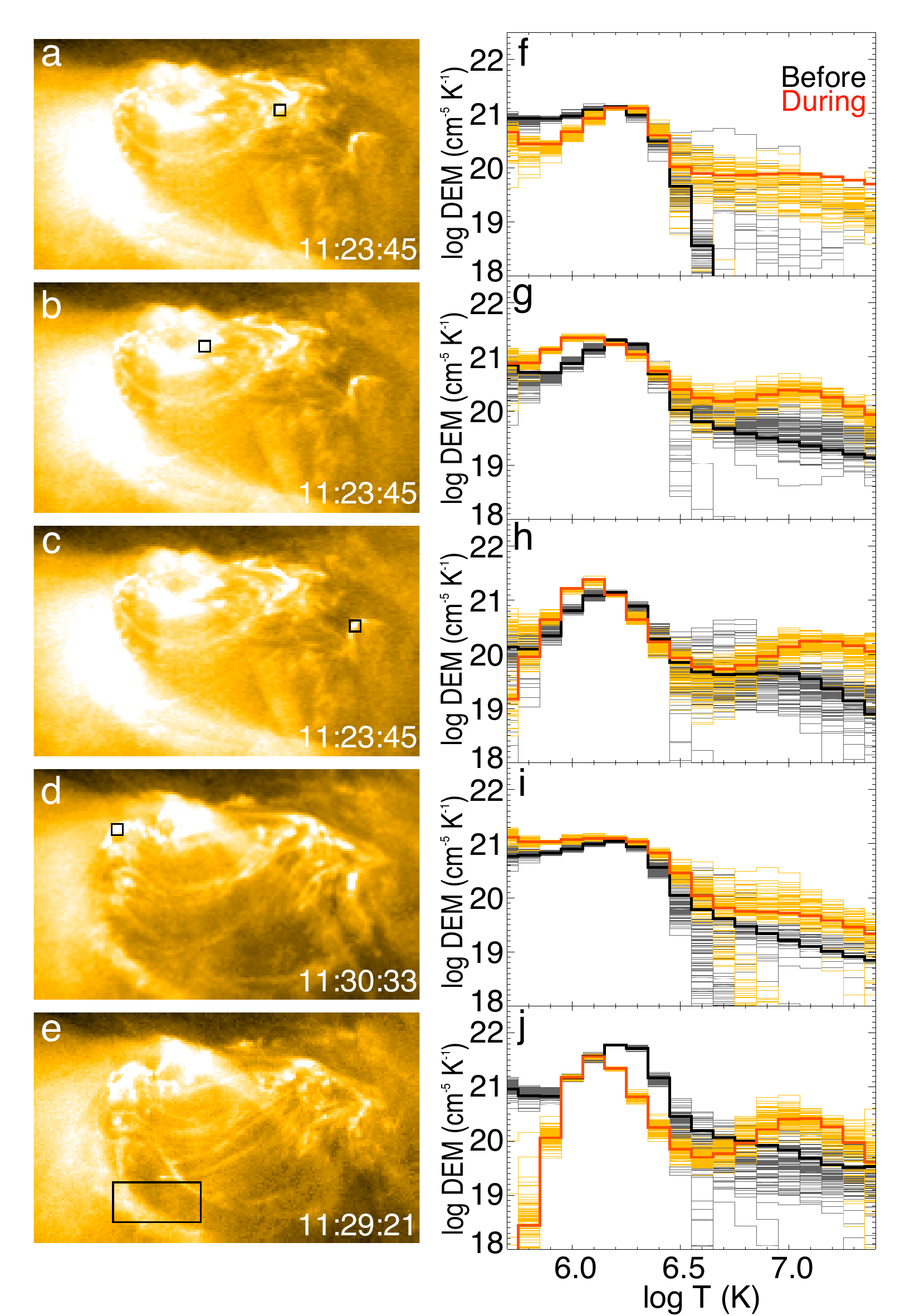}
    \caption{Thermal properties of the mini-filament eruption. Left: EUI HRI$_{EUV}$ 174 \AA\ brightenings that appear during the mini-filament eruption. Their representative positions are marked by black boxes that include flare ribbons (a-b), mini-filament footpoints (c), footpoints of the ambient loops (d), and the interaction region between the erupting mini-filament and nearby ambient loops (e). Right: DEM curves of the regions highlighted in the left panels before (10:48:45 UT; black) and during (red) the eruption. Gray and orange lines show the 100 MC simulations.} 
    \label{fig2}
\end{figure}

\begin{figure*}
    \centering
    \includegraphics[width=0.85\textwidth,clip,trim=0cm 0cm 0cm 0cm]{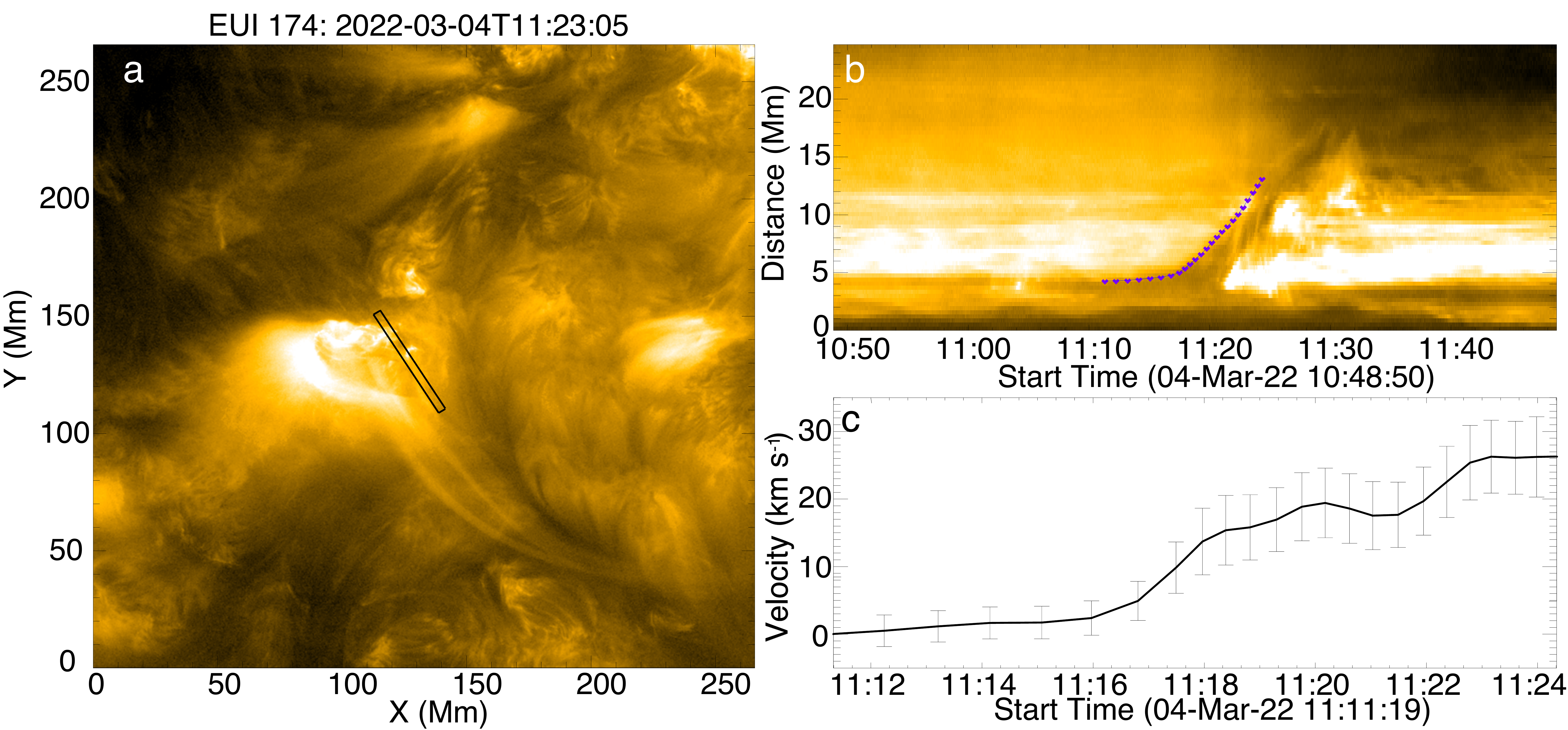}
    \caption{Kinematics of the mini-filament eruption. a. EUI HRI$_{EUV}$ 174 \AA\ image at 11:23:05 UT with a slit along the direction of the mini-filament eruption (black rectangle). b. Time-distance plot of the 174 \AA\ original images. The purple symbols denote the distance-time measurements of the mini-filament eruption. c. Temporal evolution of the velocity. The vertical bar represents its error.}
    \label{fig3}
\end{figure*}

\begin{figure}
    \centering
    \includegraphics[width=0.50\textwidth,clip,trim=0cm 0cm 0cm 0cm]{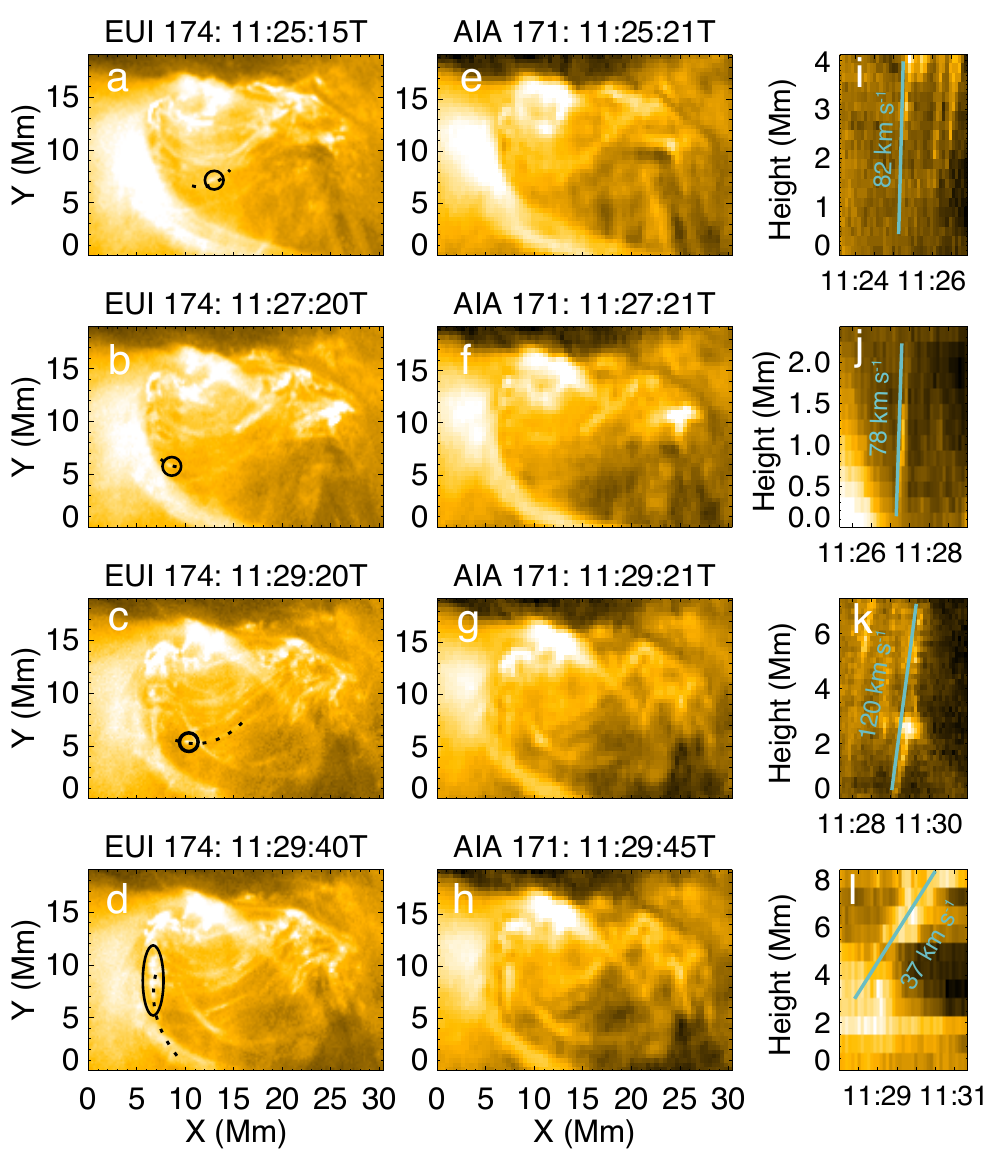}
    \caption{Details of the blobs. a-d: EUI HRI$_{EUV}$ 174 \AA\ images. The appearance of four blobs is indicated by black ellipses. The dashed black curves delineate their trajectories. e-h: AIA 171 \AA\ images at nearly the same time as a-d. i-l: Time-distance plots of the EUI HRI$_{EUV}$ 174 \AA\ images showing the fast-moving blobs corresponding to a-d.}
    \label{fig4}
\end{figure}

\section{Results}
\label{results}

\subsection{Overview of the mini-filament eruption}

The mini-filament eruption took place on 2022 March 4 and lasted for about 20 minutes, as revealed by the EUI/HRI$_{EUV}$ 174 \AA\ and AIA images in Fig. \ref{fig1}. The mini-filament with a length of $\sim$ 30 Mm presents an S-shape and is located below the far-reaching ambient loops before the eruption (green arrows in Fig. \ref{fig1}a1). This length is a typical value for mini-filaments \citep{Wang2000}. The mini-filament started to rise slowly toward the southwest at around 11:12 UT and evolved into an arch structure at about 11:23 UT, accompanied by weak brightenings of ambient loops (Fig. \ref{fig1}a2). Afterward, the mini-filament ascended quickly until its dark materials entirely dissolved into the ambient loops at $\sim$ 11:29:30 UT. Meanwhile, the width of the ambient loops increased significantly and reached the maximum, very similar to the properties of the outer spires of blowout jets (Fig. \ref{fig1}a3; \citealt{Moore2010}). After the mini-filament blew out completely, only bright post-eruption loops were visible. The ambient loops were hard to discern, possibly owing to reduced emission when they are mixed with the erupted filament plasma. 

The high-resolution EUI HRI$_{EUV}$ 174 \AA\ images reveal many more details of the mini-filament eruption, as clearly shown in zoom-in images (Fig. \ref{fig1}b). Well after the mini-filament eruption, a pair of quasi-parallel bright ribbons appeared beneath it and exhibited a separation motion (see the animation associated with Fig. \ref{fig1}). These ribbon-like structures are similar to the typical characteristics of eruptive flares caused by larger-scale eruptions, that is, they represent two rows of footpoints of flare loops that are heated by magnetic reconnection below the erupting flux \citep{Priest2002}. The bright ribbons appeared about 1.5 minutes before the obvious widening of jet spire. In addition, the west footpoint of the mini-filament (purple arrow) also became bright in the meanwhile. This indicates that part of the energy propagated there. Interestingly, a blob (in the purple ellipse) appeared in the interaction region between the erupting mini-filament and the ambient loops. The blob is explained in detail in Sect. \ref{blobs}. Furthermore, the footpoints of the ambient loops gradually brightened and are visible as a semi-circular shape (dark blue arrow in Fig. \ref{fig1}b4). This type of brightening might be related to magnetic reconnection between the erupting mini-filament and the ambient loops \citep{Aulanier2019}. The comparison of the EUI HRI$_{EUV}$ 174 \AA\ and AIA 171 \AA\ images shows that although the properties of the mini-filament eruption are similar, the first image reveals many more details of their complex structures than the second, such as two bright ribbons, brightenings at the footpoints of the mini-filament and ambient loops, and especially fast-moving blobs (Fig. \ref{fig1}c).

Although these features cannot be resolved in detail in the AIA 171 \AA\ images, the AIA images at other passbands reveal that the magnetic configuration surrounding the mini-filament appears to be quite different from the configuration that was observed in the 171 \AA\ passband before the eruption (Fig. \ref{fig1}d). The mini-filament is clearly visible at the AIA 304 \AA\ passband (green arrow in Fig. \ref{fig1}d1). In the images at the AIA 131 \AA, 211 \AA, and 171 \AA\ passbands (Fig. \ref{fig1}d2, d3, and c1), the mini-filament is not clearly identified, but is found to be enclosed by the nearby ambient loops that appear as a set of diffuse long loops extending toward the high corona. Some bright closed loops above the mini-filament also appear at the AIA 211 \AA\ passband with hotter representative temperatures (Fig. \ref{fig1}d3), indicating that in the pre-eruptive region, multiple structural components are well organized by temperature. To reveal the magnetic topology surrounding the mini-filament, we also inspected the HMI LOS magnetograms, in which the mini-filament is located above the PIL between a semi-circular flux of positive polarity and a small patch of negative polarity (Fig. \ref{fig1}d4). Moreover, we found weak flux cancelation near the local PIL, as discussed in Appendix \ref{app1}, which were frequently found in mini-filament eruptions \citep{Panesar2016b,Sterling2018,Mou2018}.

\begin{figure*}[!ht]
    \centering
    \includegraphics[width=0.99\textwidth,clip,trim=0cm 0cm 0cm 0cm]{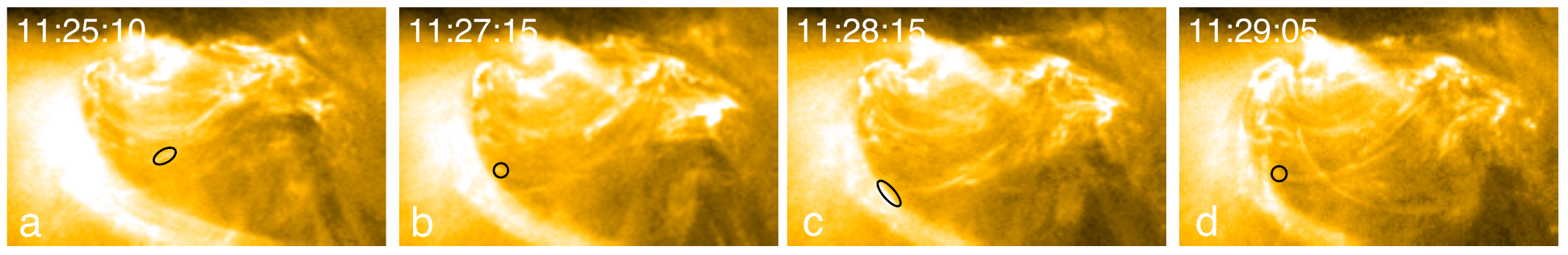}
    \caption{Locations of the blob appearance are highlighted by black ellipses.}
    \label{fig5}
\end{figure*}

\subsection{Thermodynamics of the mini-filament eruption}
The almost simultaneous multi-waveband broad-temperature observations of the AIA enable a differential emission measure (DEM) analysis \citep{Cheng2012}. We derived the DEM curves of various brightenings during the mini-filament eruption, including the bright ribbons, footpoints of the erupting mini-filament, footpoints of the ambient loops, and the interaction region of the mini-filament and the ambient loops (Fig. \ref{fig2}). In addition, the uncertainties of DEM solutions were estimated using 100 Monte Carlo (MC) simulations. The DEMs are well resolved in the temperature range from log T=5.8 to log T=7.3 given the convergence of 100 MC solutions. At 10:48 UT, before the eruption, these regions were primarily dominated by plasma with lower temperatures (5.8 $\le$ log T $\le$ 6.4). However, during the eruption of the mini-filament, plasma with higher temperature (6.6 $\le$ log T $\le$ 7.3) becomes remarkably clear in the DEM distributions in these regions. Except for the low-temperature peak of about 1.2 MK, another high-temperature component of about 10 MK exists as well. We also note that the emission measure (EM) at the lower temperature component between log T=5.8 and log T=6.4 slightly decreases, or its peak temperature shifts toward lower-temperature values than before the eruption. The precise interaction region between the mini-filament and the ambient loops is difficult to determine from current 2D images. To avoid this uncertainty, we selected the much larger region (large black box in Fig. \ref{fig2}e) in which the blobs first emerged to calculate their average DEM curves. Figure \ref{fig2}j displays one example. These results, especially the enhancement of high-temperature plasma emission, suggest that these brightenings of interest are most likely heated by the reconnection. The brightenings that appear as two ribbons below the erupting mini-filament suggest that internal reconnection takes place below the erupting mini-filament, while the brightenings distributed outside of the two ribbons might be caused by external reconnection between the mini-filament and the ambient loops.

To investigate the kinematics of the erupting mini-filament, we created a time-distance plot (Fig. \ref{fig3}b) based on the slit along the eruption direction (Fig. \ref{fig3}a). By manually measuring the positions of the leading front of the erupting mini-filament in the time-distance plot, we calculated the projected velocities of the erupting mini-filament (Fig. \ref{fig3}c). The erupting mini-filament becomes invisible when it reaches the upper corona at about 11:24:30 UT, and the velocities after this are therefore not available. The temporal evolution of the velocity shows that the mini-filament eruption presents a slow rise phase followed by a gradual acceleration phase. The velocity in the slow rise phase is maintained at about 2 km s$^{-1}$. Entering into the eruption phase, the velocity gradually increases from 2 km s$^{-1}$ at 11:16 UT to 15 km s$^{-1}$ at 11:18 UT, corresponding to an average acceleration of about 100 m s$^{-2}$. This evolution pattern is very similar to that of the large-scale flux rope eruptions \citep{Zhang2012,Cheng2020} even though the magnitudes in velocity are different. It shows that the kinematic evolution of small-scale eruptions, most likely, mini-flux rope eruptions, is similar to that of large-scale eruptions. This suggests the same eruption mechanism, independent of spatial scales.

\begin{figure*}[!ht]
    \centering
    \includegraphics[width=0.85\textwidth,clip,trim=0cm 0cm 0cm 0cm]{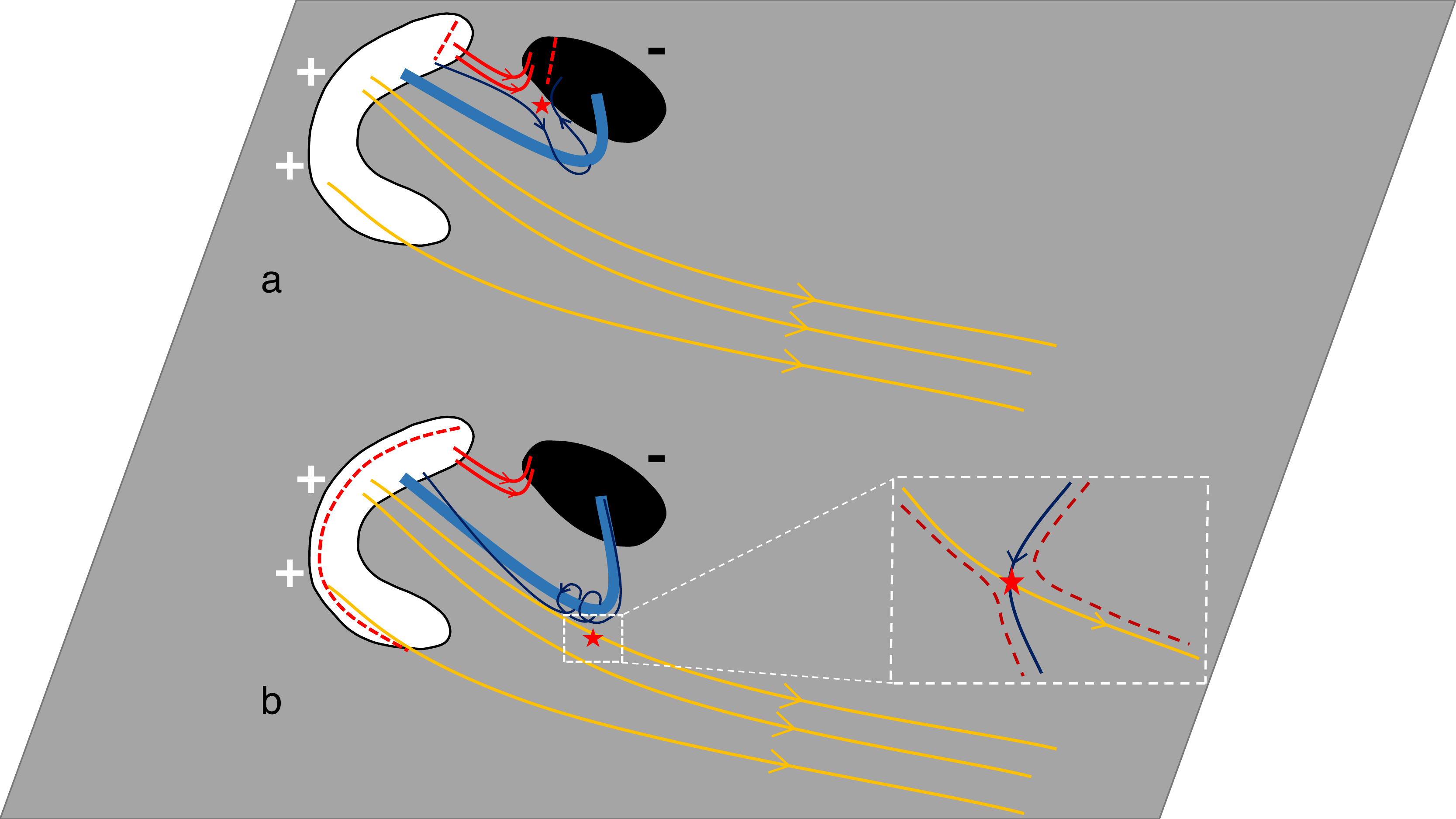}
    \caption{Schematic diagrams of the mini-filament eruption in which internal reconnection (a) first occurs, followed by the external reconnection (b). The white and black blocks represent positive and negative polarities, respectively. The long yellow lines indicate nearby ambient field lines. The thick curve in light blue denotes the erupting mini-filament, and the thin lines in dark blue surrounding the mini-filament indicate the field line above (top) and the winding of the flux rope (bottom). The red stars mark the reconnection sites. The red lines show the post-flare loops, and the dashed lines indicate the ribbons. In the zoom-in panel, the dashed red lines show newly formed field lines.}
    \label{fig6}
\end{figure*}

\subsection{Blobs during the mini-filament eruption}
\label{blobs}
The most important finding is the discovery of a chain of fast-moving blobs that are intermittently expelled from the interaction region between the erupting mini-filament and ambient loops. The attached HRI$_{EUV}$ 174 \AA\ movie clearly shows that multiple plasma blobs with a width of about 6 pixels (1.2 Mm) first appear in the interaction region and propagate along the post-eruption loops toward the footpoints of the mini-filament and ambient loops (Fig. \ref{fig4}a-d). We also show an example of a blob movement in Appendix \ref{app2}. Upon a closer examination, we find that these blobs are clearest in the 174 \AA\ images. A few blobs are also visible in the AIA 171 \AA\ images, but they cannot be fully resolved (Fig. \ref{fig4}e-h). Only the last of the four blobs shown in the 174 \AA\ images (Fig. \ref{fig4}d) can be detected in 171 \AA\ images (Fig. \ref{fig4}h), but it is vague. This highlights that a better spatial resolution, at least higher than that of AIA, is necessary to disclose these fine structures. The moving plasma blobs are not visible at other AIA passbands, possibly because their sensitivity is lower than that of HRI$_{EUV}$. Another reason might be that the blob temperature is only about the peak temperature of the 171 \AA\ and 174 \AA\ passbands ($\sim$ 1 MK). We also made time-distance plots to track the trajectories of four of these blobs (dashed black lines in Fig. \ref{fig4}a-d). We note that, the first three blobs move toward the west footpoint of the mini-filament, and the last blob moves in the opposite direction. We also estimated their linear velocities and lifetimes. The velocity of most of the blobs is about 100 km s$^{-1}$ , except for the last blob, which has a speed of 37 km s$^{-1}$. This particularly slow speed might be due to the projection effect because the angle between the moving direction and the line of sight is found to be too small. In addition, their lifetimes are very short: They vary from 30 s to 90 s, similar to the typical value of chromospheric blobs \citep{Singh2012}. A short lifetime like this is difficult to understand because the radiative cooling timescale is significantly longer. However, as suggested by magnetohydrodynamic (MHD) simulations \citep[e.g.,][]{Ni2021}, it is likely that these blobs are quickly dissolved into the ambient plasma that surrounds the reconnection region.

We suggest that these fast-moving blobs are considerable evidence of magnetic reconnection between the erupting mini-filament and the ambient loops. 
First of all, these blobs are moving fast toward the footpoints of the mini-filament, which supports the theory that reconnection occurs between the mini-filament and the ambient loops, as proposed for major flares \citep[see][]{Aulanier2019}. 
In addition, they appear at the time when the mini-filament and the ambient loops interact. Based on the attached animation, we roughly estimate that the interaction takes place between 11:24:30 UT and 11:30:00 UT, which is consistent with the time range in which the blobs appear.  
Furthermore, Fig. \ref{fig5} shows the locations of the first appearance of some blobs. They seem to have a tendency to follow the shift of the bright edge of the ambient loops. These features show that the erupting mini-filament experiences magnetic reconnection as it interacts with the ambient loops.

\section{Summary and discussions}
\label{discussion}

Using high spatio-temporal resolution data of the EUI/HRI$_{EUV}$ during the first orbit of the nominal mission phase of Solar Orbiter, we investigated the detailed process of a mini-filament that erupted outward and then reached the ambient loops. We suggest that two steps of reconnections might occur in sequence during this time. The main features of this process are summarized in Fig. \ref{fig6}. The first reconnection occurs underneath the erupting mini-filament and creates the outer twisted envelope (star in Fig. \ref{fig6}a). This process is the so-called internal reconnection, which also produces small flare loops that  manifest as JBPs, as observed by previous missions with lower resolution data. The footpoints of the small bright loops form two ribbons that are clearly revealed in the higher-resolution EUI/HRI$_{EUV}$ data. As the mini-filament rises and interacts with the nearby ambient loops, the external reconnection starts to take place (Fig. \ref{fig6}b). Because the newly formed envelope flux of the erupting mini-filament is highly twisted \citep{Aulanier2019}, it is strongly sheared relative to the nearby ambient loops. This enables reconnection at the outermost erupting mini-filament (star in Fig. \ref{fig6}b). This external reconnection resembles the component reconnection proposed by \citet{Chen2021}, but its location varies as the mini-filament erupts outward. The most important finding during this particular event is that these small-scale blobs that originate from the interaction region and then propagate along the post-eruption loops toward the footpoints of the mini-filament and ambient loops are detected unambiguously for the first time. This provides evidence of external reconnection during blowout jets.

\citet{Sterling2015} proposed that the small-scale filament eruption is the key component of blowout jets. This is clearly verified by our observations. These authors performed a statistical study and further predicted that a JBP corresponds to the flare region generated by internal reconnection underneath the erupting mini-filament. With high-resolution data, the bright flare ribbons that are concentrated in this small region are clearly revealed here. Another expectation of the model of Sterling and collaborators is that as the external reconnection progresses, the jet spire drifts away from the JBP. This is also confirmed by our observations \citep[also see][]{Baikie2022}. Interestingly, the current event also shows a distinct feature. The jet spire often brightens during the eruption, but is observed to become darker during the current event instead. One possibility to explain this is that the jet spire becomes hotter and thus fades out of the HRI$_{EUV}$ 174 \AA\ passband.

Our observations also show some new features that are different from previous results. First, \citet{Moore2018} suggested that a large majority of jets first experience external and then internal reconnection. However, in the current case, the two quasi-parallel bright ribbons appeared at about 11:22:30 UT before the obvious widening of the jet spire at about 11:24:00 UT, indicating that internal reconnection takes place earlier. Except for this, in the recent MHD model of spiral jets \citep{Wyper2017,Wyper2018}, the external reconnection is argued to occur between the field lines above the mini-filament and the overlying flux. However, based on magnetic field distribution at the photosphere, the orientation of the field lines above the mini-filament seem be similar to the overlying flux for the current event. In order to facilitate magnetic reconnection, we suggest that these field lines might first turn into the twisted flux with the internal reconnection \citep{Aulanier2019} and then reconnect with the overlying flux, as shown in Fig. \ref{fig6}b. One critical feature is that the field lines that reconnect are only highly sheared and are not ideally anti-parallel, as hypothesized in previous studies \cite[e.g.,][]{Panesar2022}. In a 3D radiative magnetohydrodynamic simulation, this type of shear-type reconnection pattern is found to be a common characteristic during small-scale flares \citep{Li2022}. 

The small-scale bright blobs with high speeds that appear in the interaction region and propagate along the post-eruption loops provide considerable evidence of magnetic reconnection between the erupting mini-filament and the nearby ambient loops. To further justify that the blobs are caused by magnetic reconnection, we estimated the local Alfv$\acute{e}$n speed with $v_A = \frac{B}{\sqrt{4\pi\rho}}$, where $B$ is the magnitude of the magnetic field and $\rho$ is the plasma density. When a fully ionized hydrogen atmosphere is assumed, $\rho = m_{i}n_{i}$, $n_i = n_e$. The electron number density is calculated by $n_e = \sqrt{EM/l}$, where the depth $l$ of the loop along the line of sight is assumed to be equal to its width, which was measured to be $\sim 10^8\ cm$. Thus, the value of $n_e$ is estimated to be $\sim 1 \times 10^{10} cm^{-3}$. Based on the 3D coronal magnetic field from a potential field extrapolation with the HMI radial field as the bottom boundary, we derived a magnetic field strength of B $\sim$ 5 G at the height equivalent to the length of the post-eruption loops. It corresponds to a local Alfv$\acute{e}$n speed of approximately 110 $km\ s^{-1}$.  Therefore, these blobs move at a velocity close to the Alfv$\acute{e}$n speed, similar to outflow features from the reconnection region that were observed previously \cite[e.g.,][]{Cirtain2013}. The external reconnection between the mini-filament and the ambient loops enables transfer of the filament material and magnetic flux to the ambient corona. However, the blobs here might also be the result of reconnection between the envelope field of the mini-filament and the ambient field, in which case they cannot directly transfer mass to the ambient corona, as argued by \citet{Huang2018}.

Using the EUI data, \citet{Mandal2022} also reported multiple high-speed propagating blob-like brightenings from the jet base, showing their close association with magnetic reconnection. However, the source region in which the blobs emerge seems to be hardly identified. In our case, they clearly originate in the interaction region of the erupting flux and ambient flux. The blobs might also be formed by braiding within the rising mini-filament threads, as suggested for the relaxation of braided loops \citep[e.g.,][]{Antolin2021,Chitta2022}. \citet{Cirtain2013} even reported that these outflow blobs moved toward the footpoints of braided loops with a speed of $\sim$ 100 $km\ s^{-1}$, which is similar to the blobs we observed here. Magnetic braiding might also lead to an increase in the loop width \cite[e.g.,][]{Schrijver2007}.

We suspect that the blobs are closely related to magnetic plasmoids that are formed in the tearing-mode instability of the reconnecting current sheet regions \citep[e.g.,][]{Li2016,Wang2021}. However, the blobs in our case are formed at varying locations and not in a particular current sheet \citep[e.g.,][]{Singh2012,Zhang2014,Ni2017,Yang2018,Zhang2019}. This indicates that the collection of various reconnection episodes in3 the space domain is also able to present a temporal evolution of magnetic reconnection in nature.

\section{Acknowledgments}
We thank the referee who helped improve the manuscript. AIA data are courtesy of NASA/SDO, which is a mission of NASA’s Living With a Star Program. Solar Orbiter is a space mission of international collaboration between ESA and NASA, operated by ESA. The EUI instrument was built by CSL, IAS, MPS, MSSL/UCL, PMOD/WRC, ROB, LCF/IO with funding from the Belgian Federal Science Policy Office (BELSPO/PRODEX PEA 4000134088, 4000112292, 4000136424, and 4000134474); the Centre National d’Etudes Spatiales (CNES); the UK Space Agency (UKSA); the Bundesministerium für Wirtschaft und Energie (BMWi) through the Deutsches Zentrum für Luft- und Raumfahrt (DLR); and the Swiss Space Office (SSO). Z.F.L., X.C., and M.D.D. are funded by NSFC grants 11722325, 11733003, 11790303, 11790300, and by National Key R\&D Program of China under grant 2021YFA1600504. L.P.C. gratefully acknowledges funding by the European Union (ERC, ORIGIN, 101039844). Views and opinions expressed are however those of the author(s) only and do not necessarily reflect those of the European Union or the European Research Council. Neither the European Union nor the granting authority can be held responsible for them. D.M.L. is grateful to the Science Technology and Facilities Council for the award of an Ernest Rutherford Fellowship (ST/R003246/1).
 
\bibliography{bibtex}
\bibliographystyle{aa}

\begin{appendix}
\section{Small-scale magnetic flux cancelation\label{app1}}
\begin{figure}[h]
    \begin{center}
    \includegraphics[width=0.5\textwidth,clip,trim=0cm 0cm 0cm 0cm]{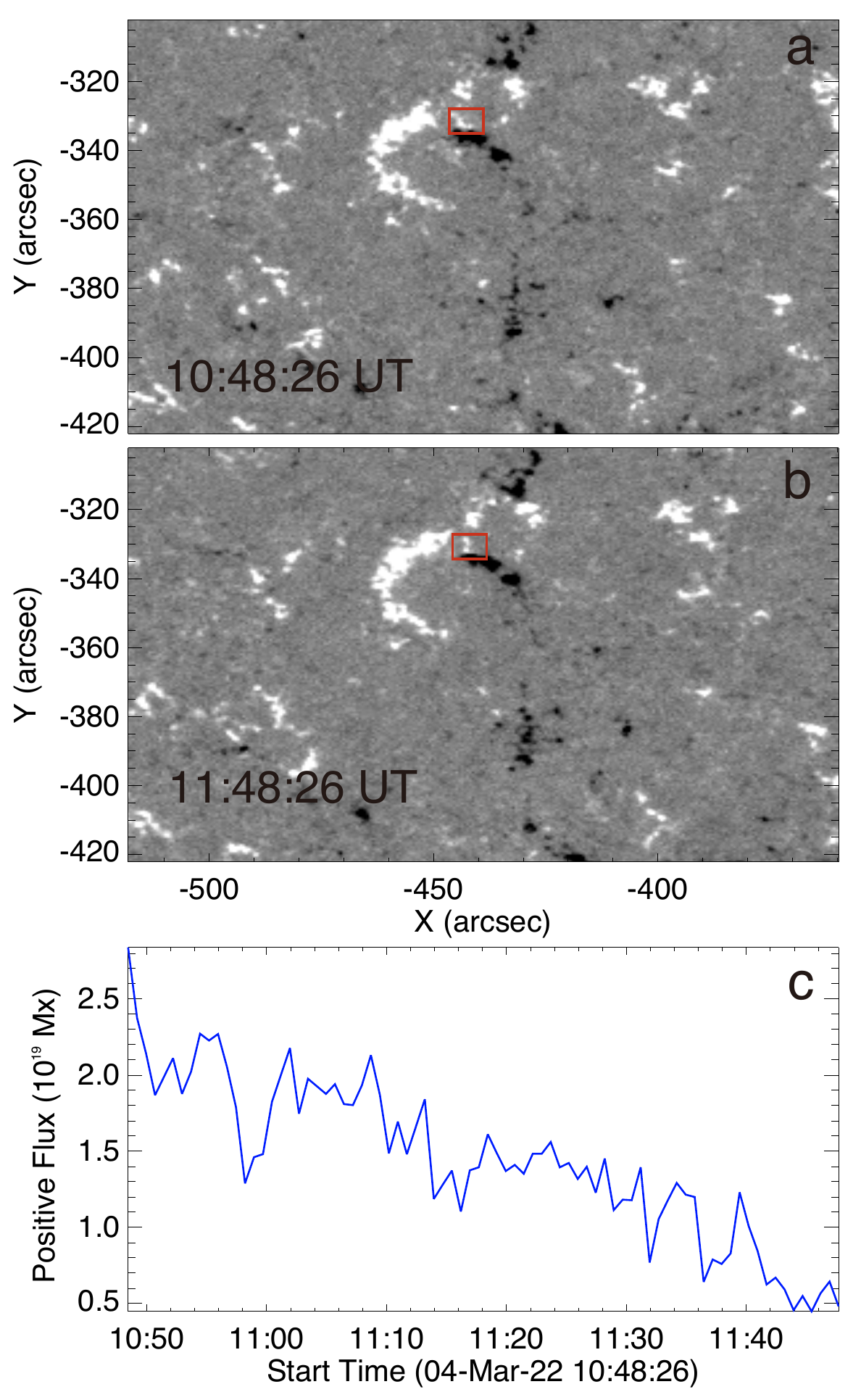}
    \caption{Magnetograms before and after the eruption on 10:48:26 UT and 11:48:26 UT are shown in panels a and b, respectively. The red boxes denote the area we used to study the temporal evolution of the flux of the minor positive polarity shown in panel c.}
    \label{figa1}
    \end{center}
\end{figure}

We found that a small positive polarity (red boxes in Fig. \ref{figa1}a and b) was mostly canceled after the mini-filament eruption. This is also indicated by the continuous decrease in positive flux, as shown in Fig. \ref{figa1}c. This is clear evidence of flux cancelation at the neutral line pertinent to the mini-filament.

\section{Example of a blob movement\label{app2}}

\begin{figure}[h]
    \begin{center}
    \includegraphics[width=0.5\textwidth,clip,trim=0cm 0cm 0cm 0cm]{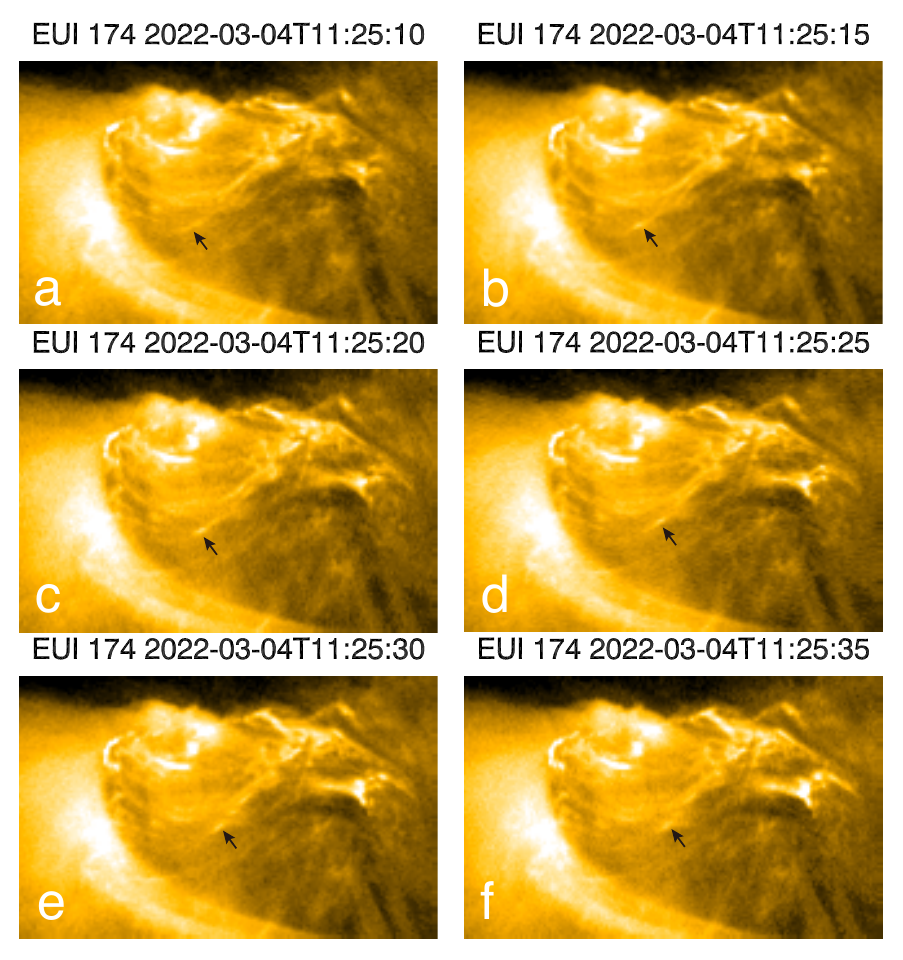}
    \caption{Tracking the motions of a blob. Its positions are pointed out by black arrows.}
    \label{figa2}
    \end{center}
\end{figure}

Figure \ref{figa2} shows an example of one moving blob that originates from the intersection region between the erupting mini-filament and the ambient loops and moves toward the footpoints of the mini-filament.

\end{appendix}

\end{document}